# Unidirectional-like Edge Transport Induced by Non-Hermitian Skin Effects

Junyi Rao[1], Jinting Ding[1], Qinghua Guo[2], Jun He[1], and Xiang Ni[1]

[1]School of Physics, Central South University, Changsha 410083, China

[2]School of Physics and Electronics, Hunan University, Changsha, 410082 China

*Non-Hermitian skin effects (NHSEs) enable dramatic boundary accumulation of waves, yet their experimental realization typically demands engineered nonreciprocity or spatially patterned loss. Here we demonstrate theoretically and experimentally that uniform loss provides a simple and previously overlooked mechanism for enforcing unidirectional-like edge transport in photonic crystals (PhCs) that breaks time-reversal symmetry in the presence of non-chiral edge states. Using a "core–cladding" geometry where domains share identical Chern numbers but possess distinct bulk polarizations, we show that uniform loss activates NHSEs that reshape the spectral topology of edge bands, giving rise to point-gap windings that dictate a one-way propagation. Near-field measurements confirm that loss converts intrinsically bidirectional interface states into a unidirectional-like circulation around the entire domain wall, showing excellent agreement with theory. Our results establish uniform loss as a universal and structurally simple route for achieving unidirectional-like wave transport.*

Topological insulators constitute a distinct phase of matter in which insulating bulk phases coexist with conducting boundary states, a phenomenon widely explored in condensed matter systems [1,2], photonics [3-6], and phononic structures [7-9]. Among various topological insulating phases, Chern insulators (CIs) provide an exemplary model, featuring chiral edge transport protected by the Chern number of the bulk system [10-12]. In photonics, the initial realization of chiral edge states in microwave frequencies [13] was soon extended into the optical regime using synthetic gauge fields [14,15]. Recent theoretical [16-18] and experimental studies[19-22], however, have revealed that even in the absence of a nonzero Chern number, a finite difference in bulk polarization might give rise to edge states and higher-order topological states. This insight substantially broadens the landscape of boundary phenomena in topological systems.

Non-Hermitian physics [23-25] has recently emerged as a powerful framework for describing systems with external energy exchanges, featuring complex energies and non-orthogonal eigenstates. Among its many research frontiers, the non-Hermitian skin effects (NHSEs) [26-42] have attracted significant theoretical and experimental interest. The NHSEs typically arise from nonreciprocal couplings in the tight-binding models [43-45], realized through active elements such as operational amplifiers in electric circuits [36,46] or directional acoustic amplifiers in phononic crystals [47]. Though effective, these active implementations complicate device architecture and risk instability when energy injection exceeds dissipation. To circumvent these limitations, passive strategies have been proposed, relying on non-Hermitian topological models with loss-induced nonreciprocal coupling, achieved through complex resonant designs[48,49], phase modulation [50], imaginary coupling engineering [51], or biased-loss configurations in resonator waveguides [52]. These works motivate further exploration of more practical routes for experimental realization of NHSEs based on controlled loss. For example, researchers have recently investigated the NHSEs through precise on-site loss engineering, and proposed the localization of chiral edge states in gyromagnetic photonic crystals (PhCs) [53-55]. However, these approaches still require highly intricate spatial loss configurations, posing significant barriers to the practical implementation of NHSEs in real devices.

In this work, we report a simple strategy to achieve unidirectional-like edge-state propagation induced by the NHSEs in lossy PhCs. By introducing a global and uniform material loss in PhCs without requiring any spatially tailored loss pattern, we experimentally realize NHSEs for edge states and subsequently achieve a transition from bidirectional to unidirectional-like edge transport, albeit without topological protection (see Fig. 1). When the two insulators are both trivial in Chern number but possess different bulk polarizations, the conventional bulk-boundary correspondence predicts the absence of chiral edge states while allowing bidirectional, time-reversal related edge states that lack topological protection against backscattering [Fig. 1(a)]. When a uniform magnetic field is subsequently applied to both regions, the two edge states that were degenerate in the zero-field case become energy-split, differentiating the CW and CCW propagating channels [Fig. 1(b)].

Crucially, introducing uniform loss opens point gaps in the complex spectrum of the edge states, each characterized by a nontrivial winding number ($\nu_{\text{I-IV}}$) associated with the corresponding interface. We show that the point-gap winding determines which edge channel is suppressed: CCW modes acquire higher loss when $v_{\text{I,III}} > 0$ $(v_{\text{II,IV}} < 0)$, whereas CW modes are attenuated when $v_{\text{I,III}} < 0$ $(v_{\text{II,IV}} > 0)$ [56]. Over a sufficient evolution period, the high-loss edge channels are selectively suppressed, leaving only the low-loss edge channel to dominate the system's dynamics. This differential dissipation effectively filters the edge-state responses, yielding unidirectional-like edge transport [Fig. 1(c)]. Unlike the conventional hybrid NHSEs of topological boundary modes, which induces corner and edge localization [53,55,57-61], the NHSEs in our 2D system selectively bias propagation direction without causing corner localization. This uniform-loss–induced directional selectivity provides a practical route to control wave transport without altering the lattice geometry, opening a new pathway toward non-Hermitian and reconfigurable topological photonics.

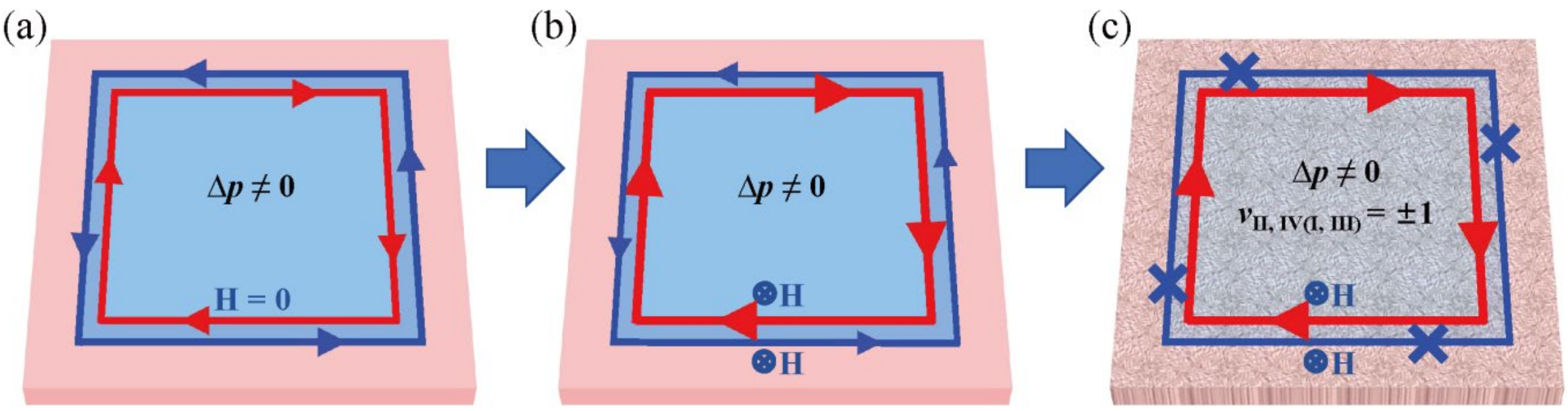

Fig. 1. Transition from bi-directional to unidirectional-like interface states via global and uniform loss. (a) With a nonzero polarization difference ($\Delta P \neq 0$), the interface supports non-chiral interface states. (b) A uniform magnetic field lifts the band degeneracy, creating an energy splitting between counter-propagating interface states. (c) Introducing global and uniform loss activates the non-Hermitian skin effect, converting interface states into unidirectional-like transport.

To verify the concept illustrated in Fig. 1(c), we propose a modified 2D Su-Schrieffer-Heeger (SSH) tight-binding model (TBM) featuring nontrivial bulk polarization. To introduce non-Hermitian effects while preserving spatial inversion symmetry, an additional lossy site is embedded within each unit cell and couples isotropically with other sites, providing a uniform background loss. A nonzero magnetic flux is introduced to break the time-reversal symmetry of the system. A schematic of the lattice structure is depicted in Fig. 2(a). The Hamiltonian is given by:

$$H(k_x,k_y)=\begin{pmatrix} & & \gamma e^{i\phi}+\lambda e^{-i\phi-ik_y} & \gamma+\lambda e^{-ik_x} & t \\ & & \gamma+\lambda e^{ik_x} & \gamma+\lambda e^{ik_y} & t \\ \gamma e^{-i\phi}+\lambda e^{i\phi+ik_y} & \gamma+\lambda e^{-ik_x} & & & t \\ \gamma+\lambda e^{ik_x} & \gamma+\lambda e^{-ik_y} & & & t \\ t & t & t & t & i\Gamma \end{pmatrix}, \qquad (1)$$

where $\gamma$ is the intracell hopping, $\lambda$ is the intercell hopping, $t$ is the coupling between the extra site and the 2D-SSH site, $\phi$ is the global magnetic flux, and $\Gamma$ is the non-Hermitian strength.

We first calculate the band structures of an expanded supercell ($\gamma<\lambda$) under Hermitian conditions ($t=\gamma=0.3$, $\lambda=1$, $\phi=0.3\pi$, and $\Gamma=0$). For these calculations, Bloch boundary conditions are applied along the $x$-direction and open boundary conditions are applied across 20 unit cells in the $y$-direction, the results are shown in Fig. 2(b). The inverse participation ratio ($\mathrm{IPR}=\sum_i|\psi_i|^4$) is represented by a color scale to trace the localization of the non-chiral edge states. The inset of Fig. 2(b) provides a magnified view of the edge bands existing within the range $E\in[-2,-1]$, revealing that the introduction of magnetic flux lifts the degeneracy of the two edge bands.

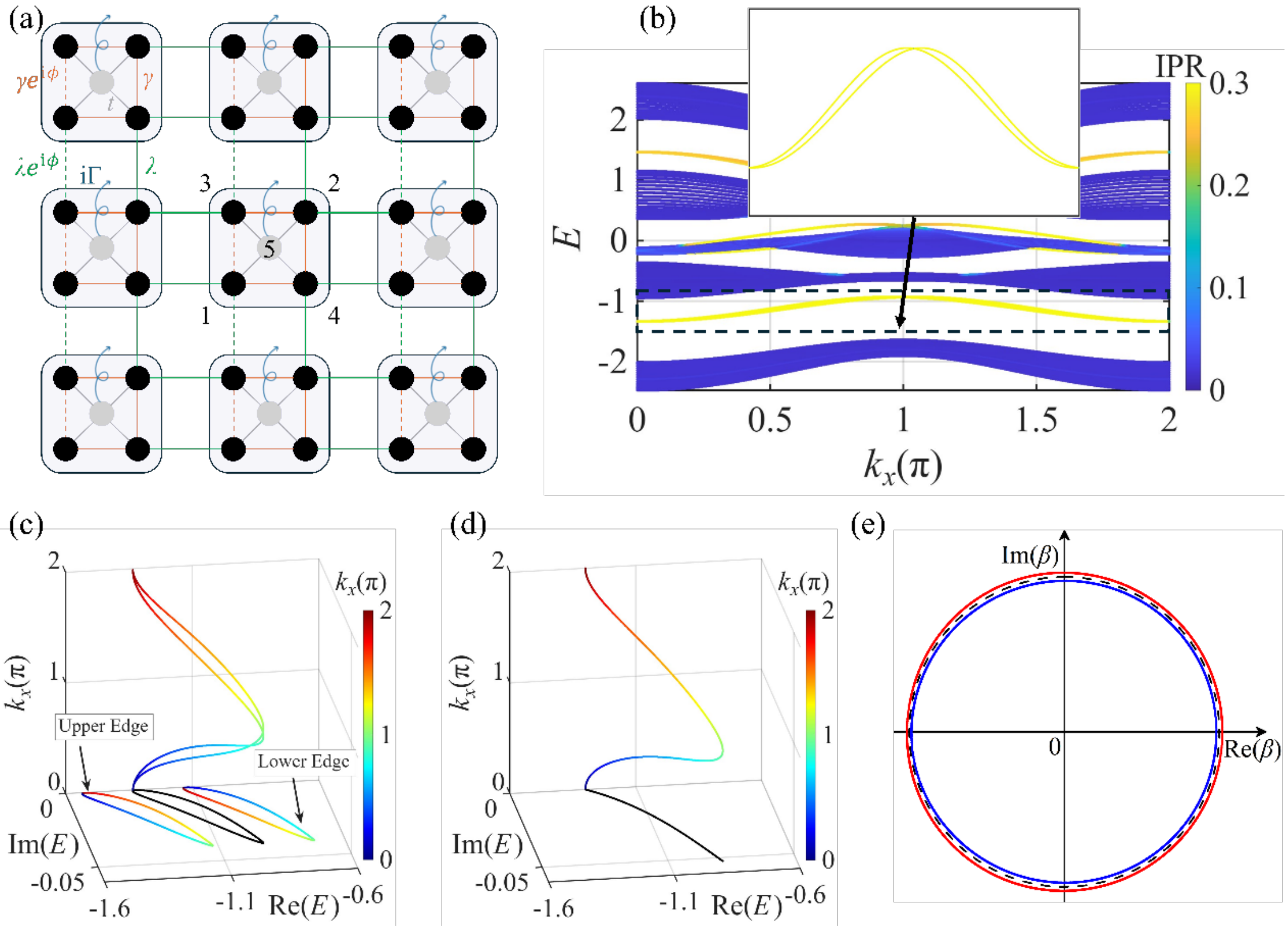

Fig. 2: Energy bands of the 2D modified SSH lattice and the generalized Brillouin zone (GBZ) of the corresponding non-chiral edge states. (a) Schematic diagram of the tight-binding model. (b) Energy bands for the Hermitian case, with parameters $t$=$\gamma$=0.3, $\lambda$=1, $\phi$=0.3$\pi$, and $\Gamma$ = 0. The inset magnifies the edge bands for better visualization. The inverse participation ratio $\mathrm{IPR}=\sum_i |\psi_i|^4$ is represented by the colormap to trace the localization of these non-chiral edge states. (c, d) Complex energy bands under non-Hermitian condition for $\Gamma$ = -0.3, with (c) $\phi$ = 0.3$\pi$ and (d) $\phi$ = 0. Both panels display the magnified view of edge bands in (b). The color mapping represents the value of $k_x$, and the bands are projected onto the complex energy plane. The upper and lower edge bands are slightly offset for clarity. (e) GBZ of edge bands. The dashed line represents the unit circle, with the red and blue contours are located outside and inside the unit circle, respectively.

To study the effects of non-Hermitian perturbation on edge states propagation, we introduce a loss of $\Gamma = -0.3$ to the auxiliary sites and consider cases both with and without magnetic flux. As shown in Fig. 2(c), when both loss and magnetic flux are present, the edge bands exhibit spectral winding in the complex energy plane. Notably, the two edge states remain degenerate in the projected plane while exhibiting opposite winding directions. In contrast, when the magnetic flux is absent, no such winding occurs in edge bands (Fig. 2(d)). This demonstrates that the emergence of spectral winding is tied to the simultaneous breaking of anomalous time-reversal symmetry and the presence of non-Hermiticity. We also numerically map the generalized Brillouin zones of the edge bands in Fig. 2(e) [56], whose trajectories lie entirely inside and outside the unit circle. Inside the unit circle, the corresponding generalized Bloch factor is less than one, producing a skin effect toward the negative direction, and vice versa. Furthermore, due to the $C_4$ symmetry of our system [56], this conclusion is equally applicable to nanoribbons along the $y$-direction. Consequently, the combination of NHSE and lattice symmetries enable the realization of unidirectional-like edge transport.

To realize the theoretical prediction of TBM in a photonic system, we design two complementary square-lattice PhCs, referred to as the "shrunken" and "expanded" configurations, respectively. (detailed in Fig. S1 of ref. [56]). We construct two supercell configurations composed of six shrunken and ten expanded unit cells, one with periodic boundary conditions (PBCs) along $y$ and open boundary conditions (OBCs) in the $x$ ("$x$-OBC/$y$-PBC"), and the other with the opposite configuration ("$x$-PBC/$y$-OBC"), neglecting material loss ( $\Gamma = 0$ ), as shown in Figs. 3(c) and 3(f). Their band structures reveal two edge states within the bulk bandgap (Figs. 3(a,b)). In the absence of a magnetic field, the two edge states are degenerate, highlighted by the violet curves in Fig. 3(a). When a uniform magnetic field is applied to both regions, this degeneracy is lifted, resulting in a clear energy splitting, highlighted by the red and blue curves in Fig. 3(b). These states localize at different interfaces: the blue branch localizes at the lower edge (labeled edge I) in Fig. 3(f) and the right edge (edge III) in Fig. 3(c), while the red branch appears at the upper edge (edge II) and the left edge (edge IV), respectively.

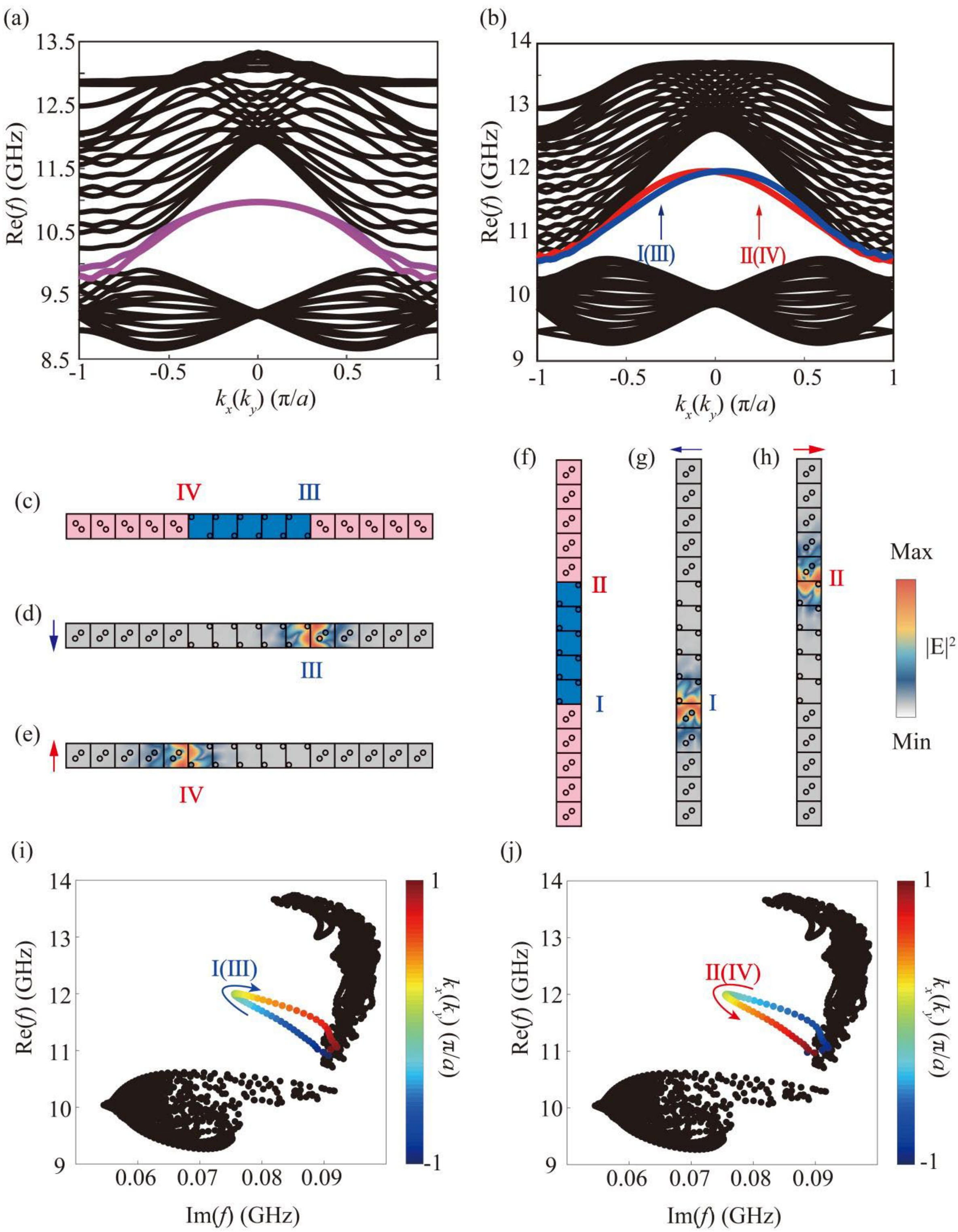

FIG. 3. Band structures of the photonic supercell and the winding behavior of non-chiral interface states. (a) Projected band structure for the supercells in (c) and (f) in the absence of a magnetic field, showing the emergence of non-chiral interface states (violet curves). (b) The corresponding band structure under a magnetic field, showing energy splitting of the interface states (red and blue

curves). (c) Supercell configuration consisting of expanded and shrunken unit cells with periodic ($y$-PBC) and open ($x$-OBC) boundaries. (d)-(e) Calculated spatial distribution of the interface eigenstates for interfaces III and IV, while $k_x = 0.25$ ($\pi/a$), with directional-selective propagation indicated by the red or blue arrows.(f) Complementary supercell with $x$-PBC and $y$-OBC.(g)-(h) Calculated spatial distribution of the interface eigenstates for interfaces I and II, while $k_y = 0.25$ ($\pi/a$), with directional-selective propagation indicated by the red or blue arrows.(i),(j) Complex band structure for supercell with $y$-PBC/$x$-OBC or $x$-PBC/$y$-OBC.To study the non-Hermitian point-gap topology of the edge states, we revisit the supercell configurations in Figs. 3(c) and 3(f), now incorporating uniform material loss ($\Gamma = -0.02$). The resulting eigenfrequencies of interface states form closed loops in the complex-frequency plane [Figs. 3(i, j)], and their spatial profiles are displayed in Fig. 3(d, e, g, h). As the Bloch wavenumber $k_x$(or $k_y$) scans the one-dimensional Brillouin zone, the corresponding eigenfrequencies of interface states trace a directional loop with distinct winding behaviors: for interface I and interface III, the loop winds clockwise. Conversely, for interface II and interface IV, it winds anticlockwise, corresponding to winding numbers $v_{\text{I}} = v_{\text{III}} = -1$, $v_{\text{II}} = v_{\text{IV}} = +1$ [56]. These opposite windings indicate the preferred propagation direction of interface states: interface states I (edge states III) favor propagation in the negative $x$ ($y$) direction, as indicated by the blue arrow in Figs. 3(d, g); whereas interface states II (interface states IV) favor the positive $x$ ($y$) direction (red arrow in Figs. 3(e, h)). As a result, the system exhibits an overall CW circulation of edge transport, effectively converting the originally bidirectional interface states into unidirectional-like ones.

To verify our theoretical prediction of the NHSEs-induced unidirectional-like interface states, we fabricated a photonic crystal prototype employing a "core-cladding" geometry, as shown in Fig. 4(a), and detailed in Fig. S2 of ref. [56]. This finite "core-cladding" structure consists of a central core of "expanded" cells surrounded by a cladding of "shrunken" cells, with absorbing boundaries applied along both the $x$ and $y$ axes. The resulting complex eigenvalue spectrum of such a structure, plotted in Fig. 4(b), reveals a set of isolated eigenfrequencies (marked by the red diamonds) lying within the bulk gaps around 12 GHz. These states exhibit no winding in the complex-frequency plane, and their spatial profiles are localized at the core-cladding interface, confirming their edge-state nature. We performed near-field microwave measurements by selectively exciting each of the four interfaces(I-IV). A source antenna was positioned near the designated interface, as illustrated in Fig. S2(a)[56], and the electric field $E_z$ was then quantitatively collected across the sample using a high-resolution near-field scanning system. Applying a spatial Fourier transform to the acquired complex field distributions along the one-dimensional paths of interfaces I-IV enabled us to reconstruct the edge bands presented in Figs. 4(c)-(f), respectively. The experimentally retrieved bands show good agreement with the corresponding theoretical edge bands [Fig. 3(b)]. Interestingly, a crucial distinction emerges: the experimental bands reveal only half of the full theoretical bands. For interfaces I and III, we observe the branches exclusively with negative group velocities [Figs. 4(d) and (f)], indicating that the waves propagate solely in the negative $x$ and $y$ directions. In contrast, interfaces II and IV exhibit only positive-velocity branches [Figs. 4(c) and (e)], confirming directional-selective propagation in the positive $x$ and $y$ directions. This pronounced spectral asymmetry provides a direct experimental signature of unidirectional-like transport, fully consistent with the point-gap winding predictions and the CW circulation of edge propagation in Figs. 3(d, e, g, h).

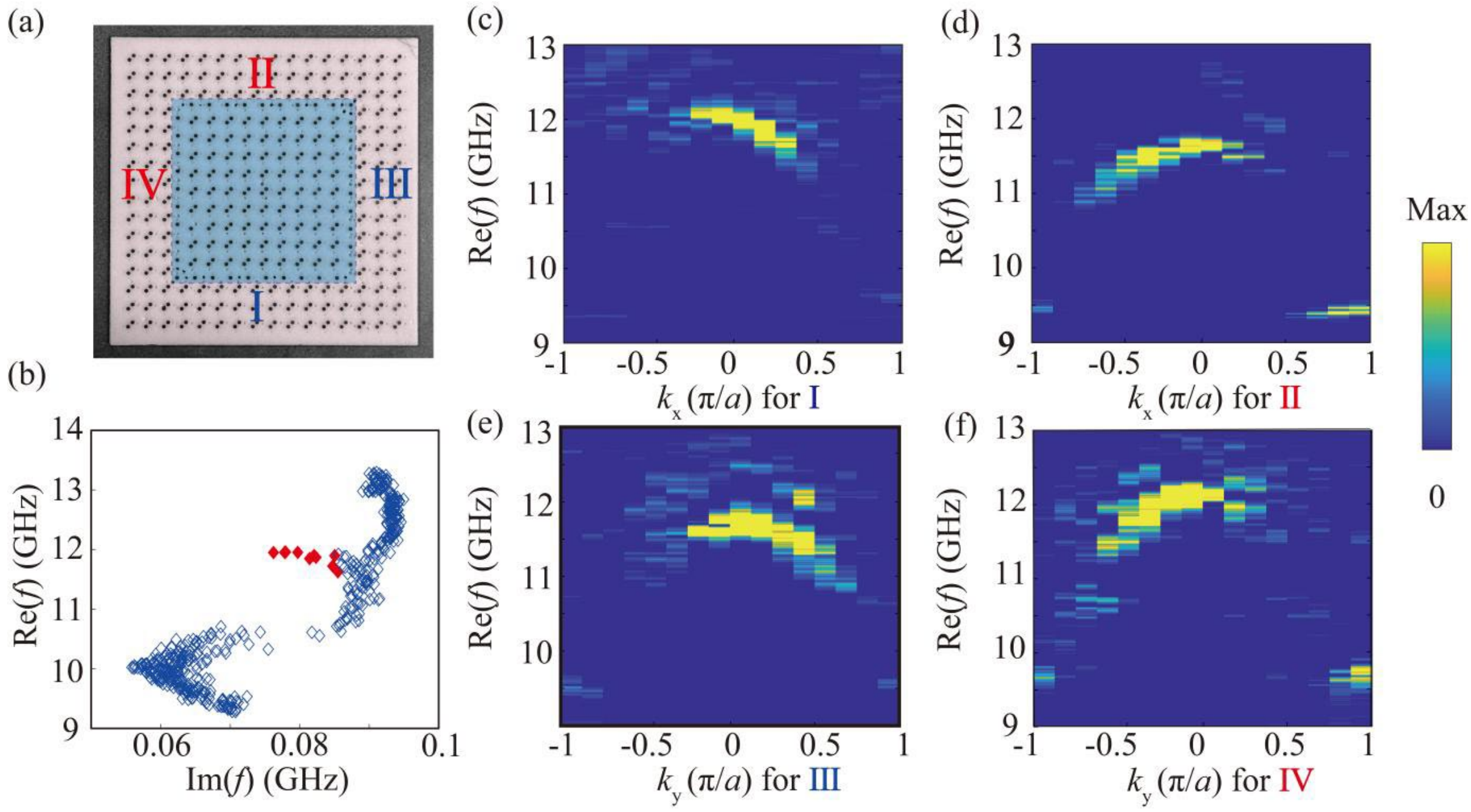

FIG. 4. The finite structure and band structure under global uniform loss. (a) Photograph of the fabricated PhC with top metallic plate removed: a central core of "expanded" cells is fully surrounded by a cladding of "shrunken" cells with absorbing boundaries applied along both the $x$ and $y$. (b) Numerically calculated complex band structure for the finite "core-cladding" geometry. (c)-(f) Experimentally retrieved interface dispersions obtained via spatial Fourier transform of the measured complex fields along interfaces I-IV.

To directly visualize this unidirectional-like circulation, we systematically moved the excitation source along the domain wall and performed near-field measurements of the electric field distribution inside the structure. The near-field images from both experiment and simulation are presented in Fig. 5, where blue pentagrams mark the respective source positions. As clearly demonstrated in Fig. 5(a), when the source is positioned on edge II, the excited waves are confined to the interface and primarily propagate in the $+x$ direction. By sequentially moving the source clockwise from edge II to III, I, and IV, we observe that the interface state consistently exhibits unidirectional-like propagation along each segment, collectively forming a closed-loop CW circulation around the entire domain wall. The remarkable agreement between the experimental results [Fig. 5(a)] and the full-wave simulations [Fig. 5(b)] further validates our theory and conclusively demonstrates how uniform material loss transforms originally bidirectional edge transport into a unidirectional-like propagation pathway.

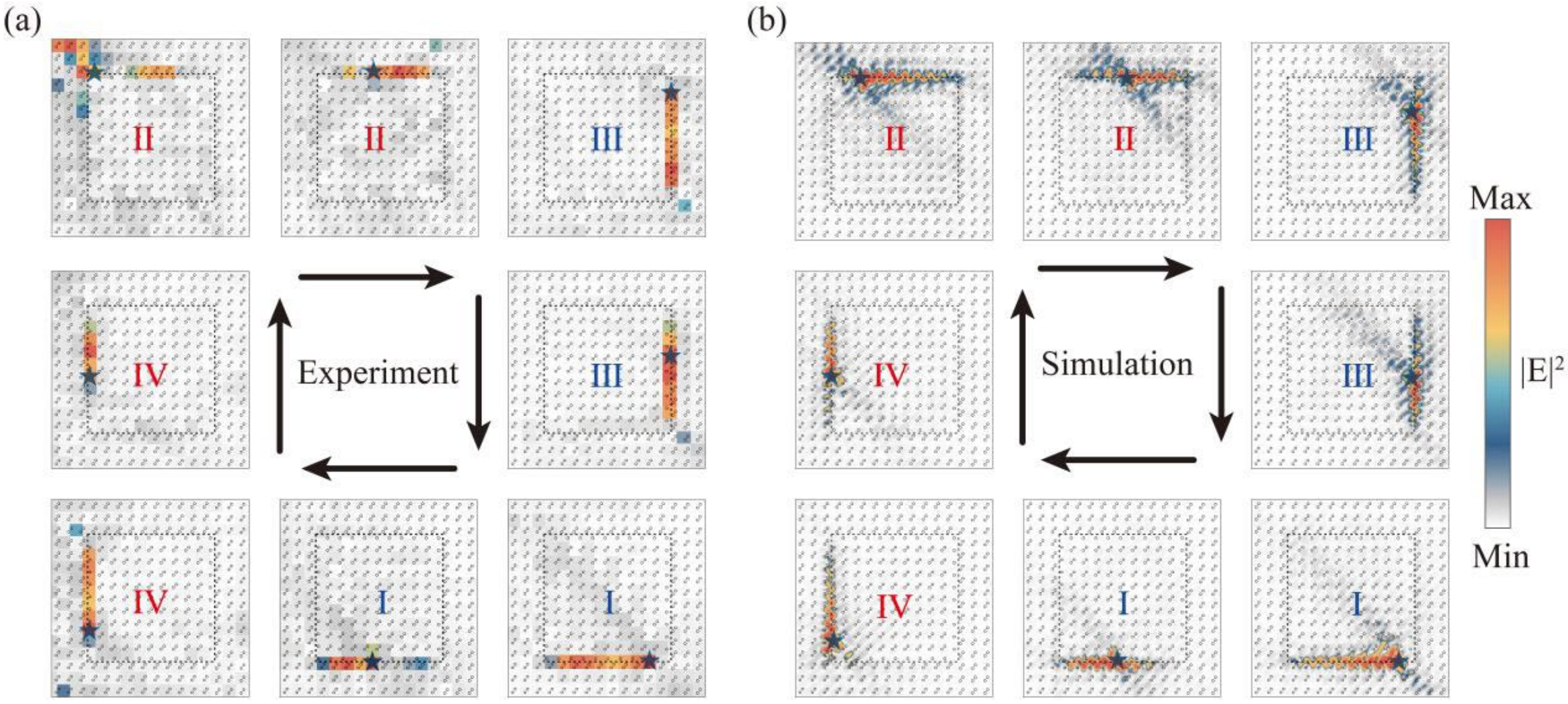

FIG. 5. Experimental (a) and simulated (b) spatial near-field imaging of unidirectional-like propagation. By sequentially moving the source antenna along the interface, the interface state exhibits unidirectional-like propagation along each segment, collectively forming a continuous clockwise circulation around the entire interface.

In conclusion, we have demonstrated that global and uniform loss in a photonic system with time-reversal-symmetry-broken but non-chiral edge states can act as an effective control knob to enforce unidirectional-like edge transport. Our approach exploits the NHSEs activated by uniform material loss to convert inherently bidirectional interface states into directional-selective ones. This transformation is rooted in the spectral topology of the lossy edge bands, whose point-gap windings dictate the preferred propagation direction. Using a simple "core–cladding" photonic architecture featuring different bulk polarizations, we have demonstrated both experimentally and through full-wave simulations that uniform loss consistently induces CW circulation of interface states along the entire domain wall. The mechanism is generic and can be applied broadly across photonic crystals and other platforms, provided that time-reversal symmetry is broken and non-Hermitian perturbation of onsite energy is introduced [56]. This paradigm of loss-enabled unidirectional-like propagation opens new opportunities for directional wave control in photonic devices. Furthermore, the circulating mode induced by NHSE in our work may have the potential to generate orbital angular momentum of waves [62].